\documentclass{emulateapj}
\usepackage{graphics}
\usepackage{graphicx}
\usepackage{color}

\usepackage{aas_macros}
\usepackage{amsmath}
\usepackage{amssymb}

\newcommand{\beq}{\begin{equation}}
\newcommand{\enq}{\end{equation}}
\newcommand{\beqa}{\begin{eqnarray}}
\newcommand{\enqa}{\end{eqnarray}}
\newcommand{\beit}{\begin{itemize}}
\newcommand{\enit}{\end{itemize}}
\newcommand{\bem}{\begin{pmatrix}}
\newcommand{\enm}{\end{pmatrix}}
\newcommand{\vecn}{\mathbf n}
\newcommand{\vecm}{\mathbf m}

\newcommand{\veck}{\mathbf{k} }

\newcommand{\lat}{\left\langle}
\newcommand{\rat}{\right\rangle}
\newcommand{\av}[1]{\lat #1 \rat}

\newcommand{\cst}{\textrm{cst}}

\newcommand{\lb}{\left [}
\newcommand{\rb}{\right ]}
\newcommand{\lp}{\left (}
\newcommand{\rp}{\right )}

\renewcommand{\max}{\mathrm{max}}
\renewcommand{\min}{\mathrm{min}}

\newcommand{\Fab}{F_{\alpha\beta}}
\renewcommand{\bem}{\begin{bmatrix}}
\renewcommand{\enm}{\end{bmatrix}}
\begin{document}

\newcommand{\muln}{\bar A}
\newcommand{\covln}{\xi_{A}}
\newcommand{\sigln}{\sigma_{\ln\rho}}
\newcommand{\murho}{\bar \rho}
\newcommand{\sigd}{\sigma_\delta}
\newcommand{\sigrho}{\sigma_{\rho}}
\newcommand{\snlnmu}[2]{s_{#1}^{\muln}(#2)}
\newcommand{\snsig}[2]{s_{#1}^{\sigln^2}(#2)}

\newcommand{\idk}{\frac{d^3k}{(2\pi)^3}}
\newcommand{\Vcell}{V_{\textrm{cell}}}.

\renewcommand{\veck}{\boldsymbol \omega}
\newcommand{\julien}[1]{\color{red} \textbf{#1} \color{black}}
\renewcommand{\cst}{\textrm{const}}
\newcommand{\kmax}{k_{\textrm{max}}}

\title{On the inadequacy of N-point correlation functions to describe nonlinear cosmological fields: explicit examples and connection to simulations}
\author{Julien Carron\altaffilmark{1,2}} 
\author{ Mark C. Neyrinck \altaffilmark{2}}

\altaffiltext{1}{Institute for Astronomy, ETHZ, CH-8093 Zurich, Switzerland}
\altaffiltext{2}{Department of Physics and Astronomy, The Johns Hopkins University, Baltimore, MD 21218, US}

\email{jcarron@phys.ethz.ch}


\begin{abstract}
Motivated by recent results on lognormal statistics showing that the moment hierarchy of a lognormal variable completely fails at capturing its information content in the large variance regime,  we discuss in this work the inadequacy of the hierarchy of correlation functions to describe a correlated lognormal field, which provides a roughly accurate description of the non-linear cosmological matter density field. We present families of fields having the same hierarchy of correlation functions than the lognormal field at all orders. This explicitly demonstrates the little studied though known fact that the correlation function hierarchy never provides a complete description of a lognormal field, and that it fails to capture information in the non-linear regime, where other simple observables are left totally unconstrained. We discuss why  perturbative, Edgeworth-like approaches to statistics in the non-linear regime, common in cosmology, can never reproduce or predict that effect, and why it is however generic for tailed fields, hinting at a breakdown of the perturbation theory based on the field fluctuations. We make a rough but successful quantitative connection to N-body simulations results, that showed that the spectrum of the log-density field carries more information than the spectrum of the field  entering the non-linear regime. \end{abstract}

\maketitle

\section{Introduction}
The non-linear regime of structure formation in the Universe is the heart of highly challenging problems for statistical inference. This regime is also potentially very rewarding, due to the large number of modes present. As seen from the point of view of statistics, the overall picture of the linear regime is very simple in principle. Since fluctuations are believed to obey Gaussian statistics at early times, an optimal description is furnished by the two point correlation, equivalently by the (power) spectrum, the second member of the hierarchy of the $n$-point correlation functions \citep{1979MNRAS.186..145W,1980lssu.book.....P,1985ApJ...289...10F,2002PhR...367....1B}. The question of optimality is however very far from clear leaving the linear regime, and clearly out of reach yet, due to our inability to model and handle accurately very high dimensional (field) statistics beyond the Gaussian. Beside the difficulties inherent in an accurate modeling of the observables on these scales, that can be approached with perturbation theory or $N$-body simulations, statistical inference also faces other types of problems. For instance, it was shown that surprisingly little information is to be extracted from the spectrum on these scales \citep{2005MNRAS.360L..82R,2006MNRAS.370L..66N,2008ApJ...686L...1L}, due to the appearance of heavy correlations between the modes. A recent approach using local transforms of the field prior the extraction of the spectrum, also applied to its weighted projection the weak lensing convergence field, was shown be successful at recapturing much information
\citep{2009ApJ...698L..90N,2011ApJ...729L..11S,2011ApJ...731..116N,2011arXiv1109.5639S,2011PhRvD..84b3523Y,2011MNRAS.tmp.1390J,2011ApJ...742...91N}. This holds at least in the absence of discreteness or shape noise issues. While it is yet not totally clear to what extent such improvements can propagate to improvements from  galaxy survey or other sort of data, it opens a new perspective on the statistics and the description of non-linear fields. The success of these transforms, and the diagonal shape of the covariance of the spectrum up to much smaller scales that it creates, suggests that a lognormal picture is not inaccurate. That is, $\ln(1 + \delta)$ may be not too far away from a Gaussian field on these scales. Some other tentative arguments for, and confirmations to some extent of this picture in lower dimensionality have been known for a long time, and in a variety of contexts \citep[e.g.]{1991MNRAS.248....1C,1995ApJ...443..479B,1996ApJ...463..409M,2000MNRAS.314...92T,2011arXiv1105.3980H}.
\newline\newline
Beside the fact that higher order correlations may carry information, another, in cosmology largely ignored process of statistical relevance is at work for tailed fields.  The correlation function hierarchy need not provide a complete description of a field anymore in this regime, so that higher order statistics may fail to capture additional pieces of information, as first pointed out in \citep{1991MNRAS.248....1C}. This possibly means that these results on the log transform of the matter field not only bring back information from higher order statistics, but also information that was lost to the hierarchy.
The one dimensional lognormal distribution is a known instance where the moment hierarchy does not specify fully its statistics. Explicit examples of other one dimensional distributions with the same moments are known \citep{Heyde63}.  For this reason, the correlation function hierarchy cannot specify fully the statistics of a lognormal field. The first quantitative evaluation of this effect, exact in one dimension, has shown that this has a huge impact on the efficiency with which cosmological parameters can be extracted from the moment hierarchy in the non-linear regime \citep{2011ApJ...738...86C}. As pointed out by \citep{1991MNRAS.248....1C}, this is a generic effect for tailed fields. Both the matter field a well as, and even more so the convergence field \citep{2006ApJ...645....1D,2011ApJ...742...15T} show large tails in the non-linear regime. Using fits to simulations, this effect was indeed shown to affect parameter inference in the one dimensional distribution of the convergence field \citep{PhysRevLett.108.071301}. Very little is however known in higher dimensional settings. It is therefore important to gain more insights on these issues, since they strongly suggest a fundamental limitation of the correlation function hierarchy in the non-linear regime.
\newline
\newline
The main purpose of this work is to make the existence of this effect within any correlated lognormal field and its correlation function hierarchy obvious. To this aim, nothing can be better than an explicit example. In section \ref{otherfields}, we will therefore present families of fields all having the same hierarchy of correlation functions at all orders as the lognormal field, for any mean and two point correlation function. We show in this light why this effect is irrelevant in the linear regime, but not in the non-linear regime. Before turning to these aspects, we discuss in section \ref{problem} why more standard approaches in cosmology, of perturbative nature, while of course perfectly sound in the weakly non-linear regime, can never predict or reproduce this effect. These are presumably reasons for which this effect has been so little studied in cosmology so far, and are worth a few comments. In section \ref{connection}, we then make a successful connection to these recent simulation results, and we conclude in section \ref{conclusion}.
The appendix collects proofs of key statements in section \ref{otherfields}.
\newline
\subsection{Notation and definitions}
We will be dealing with random vectors $\rho = (\rho_1,\cdots,\rho_d)$, being the sample a field
\beq
\rho_i = \rho(x_i) > 0.
\enq
For a vector $\vecn = \lp n_1,\cdots,n_d \rp$ of non negative integers (multiindex), we write as $\rho^\vecn$ the monomial in $d$ variables,
\beq
\rho^\vecn  = \rho(x_1)^{n_1}\cdots \rho(x_d)^{n_d}.
\enq
Throughout this work, we reserve bold letters for vectors of integers exclusively.
Let  $p_\rho(\rho)$ be a $d$-dimensional probability density function such that all correlations of the form $\av{\rho^\vecn}$ exist.  We write the moment $\av{\rho^\vecn}$ with $m_\vecn$. Explicitly
\beq
m_\vecn = \av{\rho^{n_1}(x_1)\cdots \rho^{n_d}(x_d)}.
\enq
Correlations of order $n$ are given by moments such that the order $|\vecn|$ of the multiindex, defined as
\beq
|\vecn| :=  \sum_{i = 1}^{d}n_i
\enq
is equal to $n$.
We call these quantities moments or correlations of order $n$.  These moments coincide with the values of a continuous $n$-point correlation function on the grid sampled by $(x_1,\cdots,x_d)$. We write $\delta$ for the dimensionless fluctuation field, and $A$ for the field defined by $\ln \rho$. 
\beq 
A := \ln \rho,\quad \delta := \frac{\rho -\bar\rho}{\bar \rho}.
\enq
Such assignments involving ratios or logarithms of $d$-dimensional quantities should be understood component per component. 
\section{The problem with tailed fields}\label{problem}
In one dimension, the fact that the hierarchy does not always specify fully the distribution is a well known and still active topic of research in the theory of moments in mathematics \citep[for classical references]{Shohat63,Akhiezer65,Simon97theclassical}. The moment problem is to find a distribution corresponding to a given moment series. When a unique solution exists, it is called a  determinate moment problem.
When several exist (in this case always infinitely many), it is called an indeterminate moment problem. We can refer to \citet{1991MNRAS.248....1C,2011ApJ...738...86C} for a discussion in a cosmological context and more references. The theory of the moment problem in several dimensions is less developed, but typical criteria that guarantee determinacy, or indeterminacy, linked to the decay rate of the distribution, stay basically unchanged. Guiding us throughout the discussion in this section will be the following instance: for any dimension $d$, if
\beq \label{criterium}
\av{e^{ c |\rho | }} < \infty, \quad |\rho| = \lp \rho_1^2+ \cdots + \rho_d^2\rp^{1/2}
\enq
for some $c > 0$, then the moment problem corresponding to the moments of that distribution is determinate \citep[theorem 3.1.17]{Dunkl01}.
By a  'tailed' distribution, we have in mind in this work a decay at infinity which is less than exponential, and thus for which this criterion fails. In this regime, there may thus be several distributions with the same hierarchy of correlations.
\subsection{On its relevance for parameter inference}
It should be clear why this can have in general a dramatic impact for parameter inference from correlations. Imagine a series of distributions with identical correlations at all orders, one of these distributions being the one that actually describes the observations. Since the distributions are different, they will make in general different predictions for observables other than the correlations. Pick for definiteness an observable $\av{f(\rho)}(\alpha)$ with different predictions among this family of distribution, $\alpha$ any model parameter. The knowledge of the entire hierarchy is unable to distinguish from these different predictions for $\av{f}$, since they result from equally valid distributions.
If $\alpha$ enters the true distribution in such a way that it makes a sharp prediction on the value of $\av{f}$, this is highly valuable information definitely lost to an analyst extracting correlations exclusively.
On the other hand this argument allows us also to see that this effect can become relevant only when perturbation theory breaks down. If the fluctuation field $\delta$ is small, $f$ can be expanded in powers of $\delta$, and thus $\av{f}$ can be obtained in an unique way from the correlation hierarchy of $\delta$.
\newline\newline
 There is a remarkable way to understand what is happening there in terms of Fisher information, familiar to cosmologists.  Recall that the Fisher information matrix $F_{\alpha\beta}$ associated to a probability density function $p(\rho|\alpha,\beta,\cdots)$ is defined as
\beq\label{Fab}\Fab = \av {\frac{\partial \ln p}{\partial\alpha}\frac{\partial \ln p}{\partial\beta} }.\enq
Among the many properties that makes it a meaningful measure of information are its positivity, additivity for independent variables, its invariance under invertible transformations, the Cram\'er Rao bound and the information inequality, stating that any set of observables carries at most the same amount of Fisher information as $\rho$ itself. See \citep{fisher25,Rao,vandenbos07} for references to statistical works, and \citep{1996PhRvD..54.1332J,1996PhRvL..76.1007J,Tegmark97b,1997ApJ...480...22T} for the first implementations in cosmology, for Gaussian variables. We refer to \citep{2011MNRAS.417.1938C} for an extensive discussion of the information inequality in a cosmological context, and its deep connection with the concept of entropy. It is a fact that the Fisher information content on $\alpha$ of the distribution is entirely within the first $n$ correlations if the function $\partial_\alpha \ln p$ is a polynomial of order $n$.  In particular, the distributions for which the Fisher information matrix is within the entire hierarchy are precisely those for which the functions $\partial_\alpha \ln p$ can be written as a power series over the range of $p$. If not, the mean squared residual to the best series expansion is the amount of Fisher information absent from the hierarchy \citep[for a proof]{2011ApJ...738...86C}.  It is simple to show that criterion \eqref{criterium}, that guarantees that the distribution is uniquely set by its correlations, implies as well that the entire amount of Fisher information is within the hierarchy : this follows from the very next theorem of the same reference \citep[theorem 3.1.18]{Dunkl01}, that states that the polynomials in the $d$ variables form a dense set of functions with respect to the least mean squared residual criterion, if \eqref{criterium} is met. In particular the functions $\partial_\alpha \ln p$ can be arbitrarily well approximated by polynomials with respect to that criterion,  and therefore the correlations contain all of the Fisher information. 
\newline
\newline
It is important to note that if criterion \eqref{criterium} happens to be met due to a cutoff at a large value $\rho_\textrm{cut}$, on a otherwise tailed distribution, the correlations still are poor probes for any practical purposes.
For instance, if a variable is lognormal over a very long range, but decay quickly at infinity starting from $\rho_\textrm{cut}$. Indeed, if $\rho_\textrm{cut}$ is large enough, the correlations of order up to, say, $2N$, will be identical to that of the lognormal. Since the information content of the first $N$ correlations depends on the first $2N$ only, they will be equally poor probes as for the lognormal. They will contain the exact same amount of Fisher information as the ones of the lognormal. It is the correlations of order $>N$, that are able to feel the cutoff,  that will make up for the difference between the total information content of the lognormal distribution and its correlation hierarchy (if the cutoff is at a large enough value,  from \eqref{Fab} the two distributions have the same total amount of information). The hierarchy is thus still not well suited for the analysis of data in this regime.
\newline
\newline
For the same reason, even though any lognormal field is indeterminate, this effect plays no role for parameter inference in the linear regime, when the actual range of the variables is still small, and the tail at infinity is not yet felt. This is because in this regime on one hand the lognormal is still very close to a Gaussian over the range where it takes substantial values, and thus the lowest order correlations will still contain most of the Fisher information, and on the other hand a few higher order terms are able to reproduce deviations of the functions $\partial_\alpha \ln p$ from the Gaussian very accurately over this small range. This is consistent with the findings in section \ref{otherfields} showing that the families presented there are indistinguishable form the lognormal for any practical purposes in the linear regime.
\subsection{On other approaches to non Gaussian statistics}
Let us comment in light of the criterion \eqref{criterium} on typical perturbative approaches in cosmology to parametrize (weakly) non Gaussian distributions. These involves moments, such as Gram-Charlier, Edgeworth expansions, or the relation between the moment generating function and the distribution \citep[e.g.]{1985ApJ...289...10F,1994A&A...291..697B,1994ApJ...435..536C,1995ApJ...442...39J,1995ApJ...443..479B,1998A&AS..130..193B}, in one or several dimensions. It is therefore interesting to see to what extent they fit into this picture.  Typically, when applied to the $\delta$ field, to first order these parametrize the non-Gaussianity through a polynomial with coefficients involving the cumulants, or equivalently the moments of the variable. Schematically,
 \beq \label{Edge}
 p_\nu(\nu) \propto e^{-\nu^2/2}\lp1 + \alpha_3H_3(\nu) + \alpha_4H_4(\nu) +\cdots\rp,
 \enq
 with $\nu = \delta/\sigma_\delta$. The coefficient $\alpha_i$ depends on the first $i$ moments. The correction is given in terms of Hermite polynomials $H_{n}$, which are the orthogonal polynomials associated to the Gaussian distribution. Such expansions never produce a tailed distribution, in the sense that \eqref{criterium} is always met. The decay of the distribution namely still is Gaussian. Now, to first order and over the range of $p$, equation  \eqref{Edge} is equivalent to
 \beq
 \ln p_\nu(\nu) \approx \cst - \nu^2/2 + \alpha_3H_3(\nu) + \alpha_4H_4(\nu) + \cdots 
 \enq
 Therefore, the functions $\partial_\alpha \ln p$ will have close to polynomial form.
 This is perfectly consistent with that decomposition of the Fisher information. Indeed, this expansion creates  a probability density for which its Fisher information content is within the moments that were used to build it. This is another way to see that moment-indeterminate distributions cannot be produced by perturbative expansions.

 \section{Fields with the same hierarchy of correlation functions.}\label{otherfields}
After reviewing the basic properties of correlated lognormal variables, we present both continuous as well as discrete families that have the same correlations as the lognormal at all orders, for any dimensionality $d$. In fact, it turns out that a stronger statement is true :  for these families, all observables of the form
\beq
\av{\rho(x_1)^{n_1}\cdots\rho(x_d)^{n_d}},\quad n_i = \cdots -1,0,1 \cdots
\enq
are identical to those of the lognormal field, i.e. any $n_i$ can also be negative as well. Including the hierarchy of inverse powers and 'mixed' powers to the usual hierarchy thus still does not provide a complete description.
\newline
\newline
These families are generalizations to any number of dimension, means and two-point correlations of known one dimensional examples that can be found in the statistical literature \citep{Heyde63,Stojanov87}.
\newline
\newline
Requirements such as homogeneity and isotropy are actually not needed for this section. In particular, unless otherwise specified, $\bar A$ is a $d$-dimensional mean vector $(\bar A(x_1),\cdots,\bar A(x_d))$, whose components can differ in principle. Nevertheless, the picture we have in mind is that of statistically homogeneous isotropic fields in a box of volume $V$, where some set of Fourier modes $k_\min$ to $k_\max$ can be probed. The corresponding Fourier representation of the two point correlations, in a continuous notation, is
\beq
\lb \xi_{A,\delta}\rb_{ij} = \int \idk P_{A,\delta}(k)e^{ik\cdot (x_i - x_j)} = \xi_{A,\delta}(x_i-x_j),
\enq
where the integral runs over these modes, and $\xi_{A,\delta}(r)$ is the ordinary two-point correlation function of $\delta$ or $A$. The matrix inverse is given by
\beq
\lb \xi^{-1}_{A,\delta}\rb_{ij} = \int \idk \frac{1}{P_{A,\delta}(k)}e^{ik\cdot (x_i-x_j)}.
\enq
This representation allow us to define a bit more rigorously what we mean by linear and non-linear lognormal field, or linear and non-linear regime, in the following discussion : if needed, it can be formally set as $P_{A}(k) \rightarrow 0 $ or $P_{A}(k) \rightarrow \infty $ respectively, for all $k$.
\subsection{Basic properties of lognormal fields}\label{defln}
We say the field $\rho := (\rho(x_1),\cdots,\rho(x_d))$ is lognormal if the $d$-dimensional probability density function for $A$ is Gaussian,
\beq \label{A}
p_A(A) = \frac1{\lp 2\pi |\xi_A| \rp^{d/2}}\exp\lp  -\frac 12 (A - \bar A) \cdot \xi_A^{-1} (A -\bar A) \rp,
\enq
where $\bar A$ is the mean vector of $A$, and $\xi_A$ its covariance matrix, 
\beq
\lb \xi_A \rb_{ij} = \av{\lp A(x_i) - \bar A(x_i) \rp \lp A(x_j) - \bar A(x_j) \rp}.
\enq
The probability density for the vector $\rho$ itself is then a $d$-dimensional lognormal distribution, that we define for further reference as $p^{LN}_\rho$ :
\beq \label{pLN}
\begin{split}
p^{LN}_\rho(\rho) := \frac{ p_A(\ln \rho) }{\prod_{i = 1}^d \rho(x_i)}.
\end{split}
\enq
The means and two point correlations of $A$ and $\delta$ are in one to one correspondence. We have
\beq \label{meanA}
\bar A = \ln \bar \rho - \frac 12  \sigma^2_A
\enq
where $\sigma^2_A$ is the diagonal of $\xi_A$, i.e. the variances of the individual $d$ points. Also,
\beq \label{reldA}
\lb \xi_A \rb_{ij} = \ln \lp 1 + \lb  \xi_\delta \rb_{ij} \rp,\quad \lb \xi_\delta \rb_{ij}  := \av{\delta(x_i)\delta(x_j)} .
\enq
Especially, the variances are related through
\beq
\sigma^2_A  = \ln \lp 1 + \sigma^2_\delta\rp.
\enq
\subsection{Continuous family} 
\begin{figure}
 \includegraphics[width = 8cm]{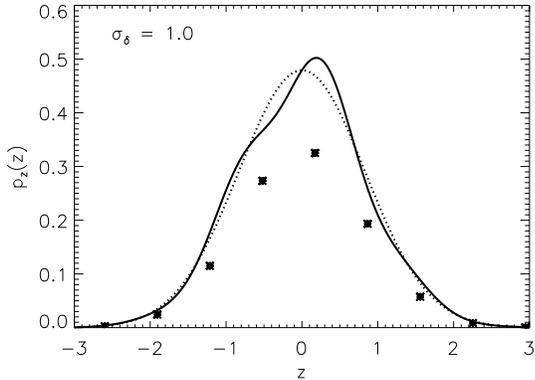}
  \caption{\label{fig:pdfs} Three different one dimensional distributions for $z = A-\bar A$, with identical moments $\av{\rho^n}, \: \rho = e^A$, for all integer $n$, positive or negative. The dashed line is the zero mean Gaussian distribution, so that $\rho$ is lognormal. The solid the member of the family in \eqref{Heydeextended} with the lowest possible frequency, and amplitude $\epsilon = 0.1$. The discrete one is \eqref{discreteP} with shift parameter $\alpha = 0.25$. They are shown at the scale of non linearity $\sigma_\delta = 1$, where this indeterminacy starts to become very relevant for inference. The families in any dimension are qualitatively identical to these. } 
 \end{figure}

\begin{figure}
 \includegraphics[width = 8cm]{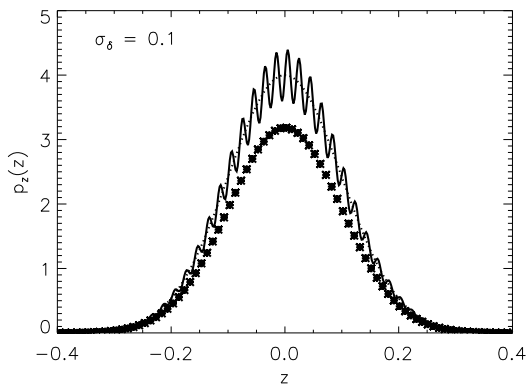}
 \caption{\label{fig:pdf2s} Same as figure \ref{fig:pdfs}, for $\sigma_\delta = 0.1$, when the indeterminacy if far less relevant for inference, for the reasons given in the text. The discrete distribution has been scaled by a constant factor for convenience.} 
 \end{figure}

Define the statistics of $\rho = \rho(x_1),\cdots,\rho(x_d))$ through the following. Pick a real number $\epsilon$ with $|\epsilon| \le 1$. Pick further a set of angular frequencies $\veck = (\omega_1,\cdots,\omega_d)$. Each of these must be an integer. Fix $p^{LN}_\rho(\rho)$ the $d$-dimensional lognormal distribution with mean $\bar A$ and covariance matrix $\xi_A$ defined above. Then set
\beq \label{Heydeextended}
p_\rho(\rho) := p^{LN}_\rho(\rho)\lb 1 + \epsilon \sin \lp  \pi \veck \cdot \xi_A^{-1} \lp A - \bar A \rp \rp \rb
\enq
Since $|\epsilon| \le 1$ this is positive and seen to be a well defined probability density function\footnote{For $d = 1$, there are very slight differences with Heyde original family. Heyde unnecessarily writes $2\pi$ instead of $\pi$, and restricts $\epsilon$ and $\omega$ to be positive.}. The claim that $p_\rho(\rho)$ defined in this way has the same moments $m_\vecn$ as the lognormal for any multiindex $\vecn$ is proved in the appendix. Note that in the above definition, $\bar A$ is the quantity that enters the definition of lognormal variables in equation \eqref{A}. It is however not the mean of $A = \ln \rho$ anymore, when $\rho$ is defined through \eqref{Heydeextended}.
\newline
\newline
The functional form of $p_\rho(\rho)$ consists of the lognormal envelope modulated by sinusoidal oscillations in $A$.  The smaller the two-point function the higher frequency the oscillations. This may sound curious at first, since it seems to imply that the more linear the field, the more different the distributions within this family will thus appear. However, this is precisely when the oscillations are the strongest that this effect is less relevant. This can be seen as the following. Taking the average of any function $f$ with respect to $p_\rho$ leads trivially to
\beq
\av{f} = \av{f}_{LN} + \epsilon \av{f  \sin \lp  \pi \veck \cdot \xi_A^{-1} \lp A - \bar A \rp \rp }_{LN},
\enq
where the subscript $_{LN}$ denotes the average with respect to the lognormal distribution.
In the limit of the very linear regime, other terms fixed, the second term will average out to zero for any reasonable $f$, since it is the integral of an highly oscillating function weighted by a smooth integrand. In the non-linear regime this in general ceases to be the case. 
This is illustrated as the solid lines in figures \ref{fig:pdfs} ($\sigma_\delta  = 1$) and \ref{fig:pdf2s} ($\sigma_\delta  = 0.1$), showing the member of that family in one dimension with minimal frequency $\omega = 1$, and $\epsilon = 0.1$. The dotted lines on these figures are the usual Gaussian for $A- \bar A = z$.  
 \newline
 \newline
The probability density function for $\ln \rho$ is not purely Gaussian anymore. It is therefore of interest to see how the correlations of $A$ deviate from those of Gaussian variables. For instance the means $\av{A-\bar A}$ do not vanish anymore as for the lognormal. A straightforward calculation leads to
\beq
\av{(A - \bar A)(x_i)} = -\epsilon \:\pi \omega_i \exp\lp-\frac{\pi^2}2 \veck\cdot \xi_A^{-1} \veck\rp.
\enq
Picking $\veck$ as having a single non zero entry, $\omega$, at $x_i$ we get that they can be as large as
\beq \label{meanz}
\begin{split}
\av{(A - \bar A)(x_i)} &= -\epsilon \:\pi \omega \exp\lp-\frac{\pi^2}2 w^2 \lb \xi_A^{-1}\rb_{ii}\rp \\
&:=  -\epsilon \:\pi \omega\exp\lp-\frac{\pi^2}2 \frac {\omega^2}{\sigma^2_{A,\textrm{eff}}(x_i)} \rp
\end{split}
\enq
Observables as simple as the means of $A$ are therefore not constrained by the knowledge of the entire correlation hierarchy of the lognormal field. While the effect is irrelevant in the linear regime (for say $\sigma_{A,\textrm{eff}} = 0.1$, the maximal value of the mean in equation \eqref{meanz} is only $\approx 10^{-215})$, deep in the non-linear regime this is not the case anymore. It is easy to show from the above expression that the range available to $ \av{\lp A - \bar A\rp(x_i)}$, choosing $w$ appropriately, scales to infinity with $\propto \sigma_{A,\textrm{eff}}$. The means are thus left totally unconstrained in that regime. This and the very sharp behavior is of course a generic effect, not limited to that particular observable. It is obvious that the relevance of this effect for parameter inference is very  sensitive to the degree of linearity of the field, and that large amounts of information are lost to the hierarchy in the high variance regime\footnote{Among this family, it turns out that some observables such as the variances $\av{(A-\bar A)^2(x_i)}$ are always identical to $\sigma^2_A(x_i)$ for any choice of $\epsilon$ and $\veck$. We do not attach any significance to this, since this is not the case for the discrete family, though closed analytical expressions cannot be obtained in this case.}.
\subsection{Discrete family}
Fix again the dimensionality $d$, the vector $\bar A$ and the matrix $\covln$. For all integer valued $d$-dimensional multiindex $\vecn$ define a realization $A_\vecn$ of $A$ as the following. Pick $\alpha = (\alpha_1,\cdots,\alpha_d)$ any point, and set
\beq \label{An}
A_\vecn := \bar A + \xi_A \cdot \lp \vecn - \alpha \rp.
\enq
While $\alpha$ can in principle be anything,  only components $\alpha_i \in [0,1)$ will actually define different grids. As usual, $\rho$ is given by exponentiation,
\beq \label{rhon}
\rho_\vecn := \exp\lp A_\vecn \rp
\enq
Assign then to these realizations parametrized by $\vecn$ a probability
\beq \label{discreteP}
P_\vecn = \frac 1 Z \exp \lp-\frac 12 \lp A_\vecn - \bar A\rp \cdot \xi_A^{-1}\lp A_\vecn - \bar A\rp\rp.
\enq
These are usual Gaussian probabilities for $A_\vecn$, except that we have only a discrete set of field realizations.
Note that it can be written, maybe more conveniently, as
\beq
P_\vecn = \frac 1 Z \exp \lp -\frac 12  \lp \vecn -  \alpha\rp \cdot \xi_A \lp \vecn - \alpha\rp\rp.
\enq
Since $\xi_A$ is positive definite, the normalization factor $Z$ is seen to be well defined, as for more usual Gaussian integrals, and so are the probabilities. This discrete probability distribution has the same moments of $\rho_\vecn$ than the $d$-dimensional lognormal distribution with associated $\bar A$ and $\xi_A$, as proven in the appendix. Again, negative entries in $\vecn$ are allowed.
\newline
\newline
This family is clearly different from the previous, continuous one. Rather than modulating the lognormal distribution with an oscillating factor, it is a series of Dirac delta functions sampling the lognormal on the grid given by \eqref{An}. The role of $\alpha$ is to shift the sample by a small amount. If $\alpha$ is set to zero, then $A = \bar A$ is part of the sample, while it is not if not. The fact that  this indeterminacy is irrelevant in the linear regime comes this time from realizing that for any nice enough function $f$, the average of $f$ will converge to $\av{f}_{LN}$ due to the trapezoidal rule of quadrature. The grid spacing at which $A$ is sampled in this way in \eqref{rhon} becomes namely thinner and thinner. In the non-linear regime, the spacing is however very large, leading again to large deviations. This is also illustrated in figure \ref{fig:pdfs} and \ref{fig:pdf2s} for the one dimensional version of it, with shift parameter $\alpha = 0.25$.
\section{Connection to simulations and discussion}\label{connection}
One of us \citep{2011ApJ...742...91N} analysed the Coyote Universe N-body simulations suite \citep{2010ApJ...715..104H,2010ApJ...713.1322L} in a box of volume $V = 2.2 $Gpc$^3$, with $256^3$ cells, extracting the spectrum  $P(k)$ of $A$ and $\delta$ over then range $0.02/$Mpc $ \lesssim k \lesssim 0.6 /$Mpc, comparing their statistical power as function of the smallest scale $k_\max$ included in the analysis for several cosmological parameters. It was found that the spectrum of $A$ has more constraining power on cosmological parameters than that of $\delta$, when the non linear scales are included in the analysis. We refer to that paper for more details on the procedures and results. In this framework, $\rho$ is $1 + \delta$, and thus $A   =\ln (1 + \delta)$. The fields are statistically homogeneous and isotropic.
\newline\newline
Given the considerations of the previous sections, and the fact that the density field is known to be somewhat  close to lognormal, these results can hardly be considered surprising. The field $A$ must be indeed closer to a Gaussian field for all values of the cosmological parameters, so that low order $N$ point functions of $A$ must contain a larger fraction of the information than those of $\delta$ (it is useful to remember that the full fields $A$ and $\delta$ carry in all cases the very same total amount of information, since the mapping between them is parameter independent and invertible). In this section we want to go a step further from these qualitative considerations and make a quantitative comparison of these results to simple analytical methods using lognormal statistics.
\subsection{Treating information in $A$ as Gaussian.}
\begin{figure}
\includegraphics[width = 0.5\textwidth]{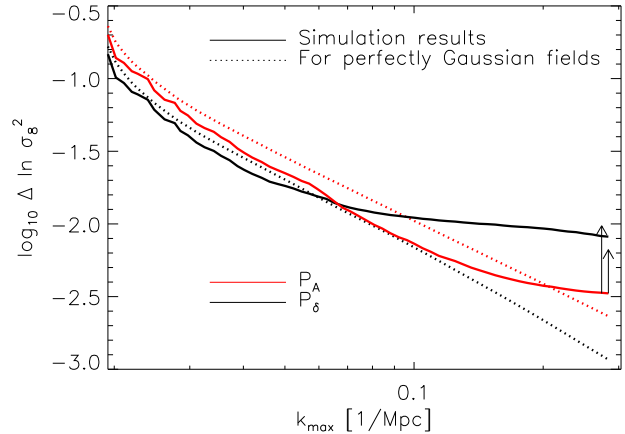}
\caption{Comparison of various estimates of the error bar on the linear power spectrum amplitude, $\ln\sigma_8^2$, constrained using power spectra of the overdensity $\delta$ (black) and the log-density $A$ (red) in an $N$-body simulation.  Solid curves show how the error bars tighten as the maximum $k$ analyzed increases up to the Nyquist frequency, as in e.g. \cite{2011ApJ...742...91N}, equation \eqref{FGnum}. Dotted curves neglect the non-Gaussian component of the covariance matrices, as well as the discrete nature of the Fourier-space mode lattice, equation \eqref{FG}.  The arrows (one for each choice of $\sigma_A$, 0.7 and 0.9) show the expected degradation of the error bars from analyzing $\delta$ instead of $A$ in our model given by equation \eqref{ratioss}; these factors appear numerically in the first column of Table \ref{table}.    \label{fig:cstr}}
\end{figure}

First, we need to make sure that a Gaussian description of the field $A$ is reasonable, at least for what concerns the information  content. In particular, this is not the case for the smallest scales of $A$, since the covariance matrix of $P_A$ in the $256^3$ box clearly shows substantial off diagonal elements starting from $k \simeq 0.3$/Mpc. We therefore repeated the same analysis, performing the logarithmic transform on the $\delta$ field only after smoothing $\delta$ on twice the original length scale, by merging the $256^3$ into $128^3$ cells. This allowed us to extract the spectra of $A$ and $\delta$ over the range  $0.02/$Mpc $ \lesssim k \lesssim 0.3 /$Mpc,  with a diagonal covariance matrix over the full range to a very good approximation. It is important to realize that sadly it is not identical to the much simpler approach of considering the original $A$ field only up to the new $k_\max$: since all the scales of $\delta$ have an impact on the large scales of $A$, the operations of smoothing $\delta$ and then log transforming $\delta$ are not identical to log transforming $\delta$ and then smoothing $A$.
\newline\newline
For a purely Gaussian field with spectrum $P$, the information content on $\alpha$ in the spectrum is given by
\beq \label{FG}
F_{\alpha\alpha}  = \frac V 2 \int \idk \lp \frac{\partial \ln P(k)}{\partial \alpha} \rp^2,
\enq
where the sum runs over the modes extracted,
and
\beq \label{cstrFG}
\frac 1 {\sqrt{F_{\alpha\alpha}}} =: \Delta(\alpha)
\enq
can be thought of as approximating the constraints on $\alpha$ achievable with these modes.  We focus for reasons that become clear below primarily on the parameter $\ln \sigma^2_8$, which has a roughly constant impact both on $\ln P_{\delta}$ and $\ln P_{A}$. In figure \ref{fig:cstr}, we compare this for the $\delta$ field and the $A$ field as function of $k_\max$. The solid lines are the simulation results, evaluating the covariance matrix $C_{kk'}$ between the modes $k$ and $k'$ and setting
\beq \label{FGnum}
\Delta(\ln \sigma^2_8) = \sum_{k,k'\le k_\max}\frac{\partial \ln P(k)}{\partial{\ln \sigma^2_8}}  C^{-1}_{kk'}\frac{\partial \ln P(k')}{\partial{\ln \sigma^2_8}},
\enq
while the dashed lines are in both cases equation \eqref{cstrFG} given by \eqref{FG}, with the derivatives being those extracted from the simulations. Since the derivatives are roughly constant, the dashed lines scale like $k^{-3/2}$, i.e. the inverse root of the number of modes. It is clear that the log transform extends the (rough) validity of the Gaussian approximation in terms of Fisher information to the full range of scales we are dealing with. Note however that this is a statement only up to the four point level, since those are the only ones that enter \eqref{FG} and \eqref{FGnum}.
\subsection{Comparison to simulations}
To compare these results to analytical predictions from lognormal statistics, we first note the following. For a parameter, such as $\ln \sigma^2_8$, that obeys roughly
\beq
\frac{\partial \ln P_A(k)}{\partial \alpha} \approx \cst = : c,
\enq
the correlated Gaussian field $A$ is equivalent, from the point of the view of the information on that parameter, to a field with the same variance but with $\xi(r) = 0$ for $r > 0$. This may not sound like an obvious statement so let us show this explicitly : start from equation \eqref{FG} which leads to
\beq
F_{\alpha\alpha} = c^2\: \frac V 2 \int \idk
\enq
The integration on the right, in a discrete description, is the number of available modes, equal to the number $d$ of grid points, times the spacing of the modes $\Delta k = (2\pi)^3/V$. It follows
\beq \label{Ftot}
F_{\alpha\alpha} = c^2 \: \frac d 2.\
\enq
On the other hand, the observation of $d$ uncorrelated Gaussian variables with variance $\sigma^2_A$ always carries the information
\beq \label{Fvar}
 d\: \lp \frac{\partial \sigma^2_A}{\partial \alpha}\rp^2 \frac 1{2\sigma^4_A}
\enq
in their variances.
If the derivative of $\ln P$ is the constant $c$, we have
\beq
\begin{split}
\frac{\partial  \sigma^2_A}{\partial \alpha} &=\int \idk  \frac{\partial P_A(k) }{\partial\alpha } \\ &= \int \idk P_A(k) \frac{\partial\ln P_A(k)}{\partial \alpha} \\& =  c
\: \sigma^2_A.
\end{split}
\enq
and thus expressions \eqref{Ftot} and \eqref{Fvar} are identical. In terms of information on such parameters, the correlated, Gaussian $A$ field is thus exactly equivalent to $d$ uncorrelated Gaussian variables with the same variances. These parameters  can be seen as entering therefore predominantly the variance, the two point correlation function at zero lag, that contains most information, and the correlations at non zero lag carrying little independent information. This is also expected to hold for the $\delta$ field, since it is very non-linear and the variance dominates  over the clustering in the two point correlation matrix, i.e. the two point correlation matrix is close to diagonal, so that the variance will dominate in any covariance matrix, as well as in the sensitivity to the parameter.
\newline\newline
Since information just adds up for any number independent variables, this means that we can try and use directly the exact results one of us derived \citep{2011ApJ...738...86C} for the one dimensional lognormal distribution to get a rough but still reasonable estimate of the improvement in the constraints from analyzing the $A$ field. In that work were derived the cumulative efficiencies
\beq
\epsilon_N^\sigma = \frac{1}{F_{\alpha\alpha}} \sum_{n = 2}^N\lp s_n^\sigma\rp^2  \in (0,1)
\enq
of the first $N$ moments of the $\delta$ field to catch the information in $A$ (equations 31-35 as well as figure 2 and figure 1, solid line, in that paper). These coefficients are extremely sensitive functions of $\sigma^2_A$, decaying like $\exp(-4\sigma^2_A) \sim \sigma_\delta^{-8} $ as soon as $\sigma_A$ becomes close to unity.
\newline\newline
There is a slight modification to make to these coefficients so that we can confront them to the simulations. From the simulations only the spectrum of $A$ were extracted, but not the mean of $A$, which also carries information in principle, even if $\delta$ itself has zero mean. For a one dimensional lognormal variable with unit mean, we have from equation \eqref{meanA} that $\bar A = -\frac 12 \sigma^2_A $. For that lognormal variable the total information is given by the usual formula for the Gaussian $A$,
\beq
F_{\alpha\alpha} = \frac{1}{\sigma^2_A}\lp \frac{\partial \bar A}{\partial \alpha} \rp^2+ \frac{1}{2\sigma^4_A}\lp\frac{\partial \sigma^2_A}{\partial \alpha}\rp^2.
\enq
It reduces thus to
 \beq
 F_{\alpha\alpha} = \frac{1}{2\sigma^4_A}\lp\frac{\partial \sigma^2_A}{\partial \alpha}\rp^2 \lp 1 + \frac{\sigma^2_A}{2}\rp,
\enq
where the rightmost term contains the part of the information in the mean of $A$.
The efficiencies ratios of the moments of $\delta$ to that of the variance of $A$ only, excluding the mean, becomes thus
\beq
\tilde \epsilon_N^\sigma:= \epsilon^\sigma_N \lp 1 + \frac {\sigma^2_A} 2\rp.
\enq
Note that in principle theses efficiencies can now be larger than unity, if the moments of $\delta$ would capture not only the information in $\sigma^2_A$, but also that in $\bar A$.
\newline
\newline
The improvement factors, i.e. the ratio of the constraints on $\alpha$ from analyzing the first $N$ correlation functions of $\delta$, to the the constraint from the two-point function of A, are thus in this model
\beq \label{ratioss}
\lb \tilde \epsilon^\sigma_N \rb^{-1/2} =: \Delta^\delta_N(\alpha)/ \Delta^A_2(\alpha).
\enq
They are independent of the parameter $\alpha$ in this one dimensional picture, since the only relevant parameter is $\sigma^2_A$, or equivalently $\sigma^2_\delta$. Remember that the denominator on the right hand side can actually be calculated for any lognormal field from \eqref{FG}, our additional assumptions can be seen thus as entering only the numerator. We argued that this ratio is expected to be correct for parameters such as $\ln \sigma^2_8$, but they become in all cases exact for a lognormal field whose variance dominates enough the clustering, $\xi_\delta(r) / \sigma_\delta^2 \ll 1$, for all $r$. The effective nearest neighbor distance given the modes we used can be evaluated as $r_\min \approx \int \idk ^{-1/3}$,  and we find $\xi_\delta(r_\min) / \sigma^2_\delta = 0.3$.
\newline\newline
Finally, there is slight ambiguity in evaluating $\tilde \epsilon_n^\sigma$. A purely lognormal field has $\sigma_A = [\ln (1+\sigma^2_\delta)]^{1/2}$, but this relation is not fulfilled precisely in our simulations. We obtain $\sigma_A = 0.7, \sigma_\delta = 1.1$ and so $, [\ln (1 + \sigma_\delta^2)]^{1/2} = 0.9$ rather than $0.7$. This discrepancy may be due of course to an intrinsic failure of the lognormal assumption, or to the presence of the smallest scales, slightly correlated, as seen from the start of saturation in figure \ref{fig:cstr}. 
\newline\newline
We show in the first two rows of Table \ref{table} the factors of improvement for these two values of $\sigma_A, 0.7 $ and $0.9$, for $N= 2,3$ and $\infty$.  In the third row is shown the improvement found extracting $P_A$ rather than $P_\delta$ in the simulations. Given our assumptions, and the very high sensitivity of $\epsilon_N^\sigma$ to the variance of the field, they agree remarkably : for the sake of comparison, a variance twice as large of $\sigma_A = 2 \rightarrow \sigma_\delta = 7.3$ would have predicted a factor of $\Delta_2 (\delta)/ \Delta_2(A) = 522$, and for $\sigma_A = 3 \rightarrow \sigma_\delta = 90$ a factor of $\approx 5\cdot10^6$. 
\newline
 \newline
We also performed this analysis for the tilt parameter $n_s$, which from its very definition has a very differentiated impact over different modes, and finding, just as in the original analysis \citep{2011ApJ...742...91N}, that the improvement factor is roughly parameter independent as shown in the fourth row of the table. This is another argument supporting the view that the dynamics of the information are indeed captured by such a simple picture. It may be due to the fact that the smallest scales, containing the largest number of modes, contributes the majority of the information in $A$ for any parameter, and thus that the sensitivity can be effectively treated as constant, equal to its value on small scales, making our argument above valid for basically any parameter. Note that for both values of $\sigma_A$ the spectrum of $A$ still outperforms the entire hierarchy of $\delta$ by a sizeable factor for the lognormal model. Of course, this is much more speculative.




\begin{deluxetable}{lccc}




\tablecaption{Factors of improvement in constraints on parameters \label{table}}

\tablenum{1}

\tablehead{\colhead{} & \colhead{$\Delta^\delta_2 / \Delta^A_2$} & \colhead{$\Delta^\delta_3 / \Delta^A_2$} & \colhead{$\Delta^\delta_\infty / \Delta^A_2$} \\  } 

\startdata
LN, $\sigma_A = 0.7$ & 2.0 & 1.6 & 1.3 \\
LN, $\sigma_A = 0.9$ & 2.9 & 2.4 & 2.1 \\
Sim. $\alpha = \ln \sigma^2_8$ & 2.5 & & \\
Sim. $\alpha =  n_s$ & 2.4 & & \\
\enddata




\end{deluxetable}


\section{Conclusion and discussion} \label{conclusion}
We have made clear that the correlation functions are generically very poor descriptors and probes of fields with large tails. This is especially true for the lognormal field, a standard prescription for the statistics of cosmological non-linear fields, and we provided other explicit fields with exactly the same hierarchy at all orders. We showed that the knowledge of the entire hierarchy of $N$ point functions of a non-linear lognormal field is insufficient to constrain other, simple, observables. We discussed the links between these aspects and the failure of power series expansions to reproduce relevant functions. We argued that this inadequacy is responsible for the recent successes of the log transforms in cosmology at recapturing information, and that they may not only bring back information from higher order statistics, but likely also information that cannot be probed at all with the hierarchy. We then showed that the factors of improvements on constraints from analyzing the spectrum of $A$ to that of $\delta$ as seen in $N$-body simulations are in quantitative agreement with simple analytical predictions using lognormal statistics.
\newline
\newline
Observational noise issues were not considered in this work. It remains therefore unclear to what extent these improvements can be achieved with actual galaxy survey data. Generically, it is reasonable to expect that noise will reduce these improvement factors. This work nonetheless makes clear that in this case, improving the specifications of a survey in order to decrease the observational  (e.g. shot) noise will be at the same time actually reducing the efficiency with which cosmological parameters can be extracted with the hierarchy of $\delta$ (i.e. the fraction of information that is contained in the hierarchy with respect to the total).
\newline
\newline
Surely, the question of the incompleteness of the hierarchy of the matter or any other field is in itself to a certain extent academical, since high order correlations will probably anyway stay out of reach for a long time. Nevertheless, it provides directions and insights into the recent successes of these transforms, strongly suggesting that in the non-linear regime, an approach using transforms is much more promising than targeting higher order statistics for inference on any parameter. We are also convinced that the statistical methods and formalism introduced will be more widely applicable in the future. Progress on these issues will be reported in due time.
\acknowledgements
We are thankful to the anonymous referee. JC warmly thanks Alex Szalay, Xin Wang and the hospitality of the Physics and Astronomy Department of Johns Hopkins University, where this work was conducted. He also acknowledges the support of the Swiss National Foundation. MCN is grateful for support from the W.M. Keck and the Gordon and Betty Moore Foundations.
\clearpage
\newpage

\bibliographystyle{apj}
\bibliography{bibMVln}

\appendix

We prove the claim made in this work that the distributions we defined have the same correlations than the lognormal at all orders.  As we will see this is also true including 'negative orders' and 'mixed orders', i.e. when negative powers of the variables are allowed in the correlations. 
\newline
Recall that for lognormal variables $\rho = (\rho_1,\cdots,\rho_d)$ with means and covariance matrix of their logarithms $\bar A = (\bar A _1,\cdots \bar A_d)$ and $\covln$ we have
\beq \label{mln}
m_\vecn := \av{\rho^\vecm}=  \av{\rho_1^{n_1}\cdots \rho_d^{n_d}} =  \exp \lp \vecn \cdot \bar A + \frac 12 \vecn\cdot \covln \vecn \rp,\quad \vecn = (n_1,\cdots,n_d).
\enq
A simple proof of this fact is to make use of the standard formulae for Gaussian integrals, valid for any positive matrix $\xi_A$, mean vector $A$ and vector $z$, that can be complex valued.
\beq \label{Gintegral}
\frac{1}{\lp 2\pi  \rp^{d/2}}\frac{1}{\sqrt {\det \xi_A}} \int d^d A \exp\lp-\frac 12\lp A - \bar A\rp \cdot \xi_A^{-1}\lp A - \bar A\rp + \lp A - \bar A \rp \cdot z \rp = \exp\lp \frac 12 z \cdot \xi_A z\rp.
\enq
Essentially all calculations in this work follow from this formula. Even the proof for the discrete family can be considered a discrete version of that relation.
\subsection{Continuous}\label{continuous}
\noindent
To prove our claim it is enough to show that
\beq \label{toprove1}
\av{\rho^\vecn \sin \lp \pi  \:\veck \cdot \covln^{-1}\lp A -  \bar A\rp\rp }_{LN} = 0.
\enq
This must hold for any  $d$-dimensional multiindices $\veck $ and $\vecn$ (we allow entries to be negative), where the average is taken with respect to the lognormal density function, equation \eqref{pLN}. We proceed as the following : we evaluate the following integral
\beq \label{A4}
I(\vecn,\veck) := \av{\rho^\vecn \exp \lp i\pi  \:\veck \cdot \covln^{-1}\lp A -  \bar A\rp\rp }_{LN},
\enq
and show that its imaginary part vanishes for $ \veck$ and $\vecn$ as specified.
\newline
Writing equation \eqref{A4} using
\beq
\rho^\vecn = \exp(\vecn \cdot A) = \exp\lb \vecn \cdot \lp A - \bar A \rp + \vecn \cdot \bar A \rb
\enq
leads immediately to the Gaussian integral given in \eqref{Gintegral}, with $z = \vecn +  i \pi \xi_A^{-1}\veck$. It follows from that equation
\beq
I(\vecn,\veck) = \exp \lb \vecn\cdot \bar A + \frac 12 \lp \vecn + i\pi \covln^{-1}\veck \rp \cdot \covln  \lp \vecn +i \pi \covln^{-1}\veck \rp  \rb.
\enq
Separating real from imaginary argument, this expression reduces to
\beq
\begin{split}
I(\vecn,\veck) =& \exp \lp  \vecn\cdot \bar A + \frac 12 \vecn \cdot \covln \vecn - \frac{ \pi^2}{2}\veck\cdot \covln^{-1}\veck   \rp \cdot \exp \lp i\pi  \: \veck \cdot \vecn \rp.
\end{split}
\enq
The imaginary part of that expression is thus proportional to $\sin \pi\: \veck \cdot \vecn$. Whenever $\veck$ and $\vecn$ are integer valued, so is their scalar product $\veck \cdot \vecn = \sum_i \omega_in_i$. Therefore, the sine vanishes and \eqref{toprove1} is proved.
\subsection{Discrete}
\noindent
From equation \eqref{An} and \eqref{rhon}, we have
\beq
\rho_\vecn^\vecm = \exp \lp \vecm \cdot \bar A + \vecm \cdot \xi_A\lp \vecn - \alpha \rp \rp.
\enq
It follows that the moments of $\rho$ are given by
\beq
\av{\rho^\vecm} = \frac {e^{\vecm\cdot \bar A}} Z \sum_{\vecn \in \mathbb Z^d} \exp \lb -\frac 12 \lp \vecn -\alpha \rp\xi_A \lp \vecn-\alpha\rp + \vecm\cdot \xi_A\lp \vecn - \alpha \rp\rb.
\enq
The proof is based on completing the square in the exponent, in perfect analogy of standard proofs of the Gaussian integral in \eqref{Gintegral}. Write
\beq
 -\frac 12  \lp \vecn -\alpha \rp \cdot \covln  \lp \vecn -\alpha \rp + \vecm \cdot \covln  \lp \vecn -\alpha \rp = -\frac 12 \lp \vecn - \vecm -\alpha \rp \covln \lp \vecn - \vecm - \alpha\rp + \frac 12 \vecm\cdot \covln \vecm,
\enq
and then perform the shift of summing index $\vecn \rightarrow \vecn + \vecm$, obtaining
\beq
\av{\rho^\vecm} = \exp \lp\vecm \cdot \bar A + \frac 12 \vecm \cdot \covln \vecm\rp \frac 1 Z \sum_{\vecn \in \mathbb Z^d}\exp\lp-\frac 12 \lp \vecn -\alpha \rp \cdot \covln \lp \vecn -\alpha \rp \rp.
\enq
Since the sum ranges over all the multiindices, the shift does not create boundary terms.
This last sum is nothing else than $Z$, so that we recover
\beq
\av{\rho^\vecm} = \exp \lp\vecm \cdot \bar A + \frac 12 \vecm \cdot \covln \vecm\rp,
\enq
which are indeed the same as the lognormal in \eqref{mln}. Again, this is also true if negative entries in $\vecm$ are permitted.
\end{document}